\documentclass[12pt]{iopart}
\usepackage[super,compress]{cite}
\usepackage{graphicx}
\usepackage{subfigure}
\expandafter\let\csname equation*\endcsname\relax
\expandafter\let\csname endequation*\endcsname\relax
\usepackage{amsmath}
\usepackage{mathtools}
\usepackage{iopams}
\usepackage{amssymb}
\usepackage{breakurl}
\usepackage{url,hyperref}
\begin{document}

\title[Cumulants of the chiral order parameter]{Cumulants of the chiral order parameter at lower RHIC energies
}
\author{C Herold$^1$, A Limphirat$^1$ P Saikham$^1$, M Nahrgang$^2$}
\address{$^1$ Center of Excellence in High Energy Physics \& Astrophysics, \\ Suranaree University of Technology, Nakhon Ratchasima 30000, Thailand}
\address{$^2$ SUBATECH UMR 6457 (IMT Atlantique, Université de Nantes, IN2P3/CNRS), 4 rue Alfred Kastler 44307, Nantes, France}

\ead{herold@g.sut.ac.th}
\vspace{10pt}

\begin{abstract}
We study cumulants of the chiral order parameter as function of beam energy as a possible signal for the presence of a critical end point and first-order phase transition in the QCD phase diagram. We model the expansion of a heavy-ion collision by a fluid dynamic expansion coupled to the explicit propagation of the chiral order parameter sigma via a Langevin equation. We evolve the medium until a parametrized freeze-out condition is met where we calculate event-by-event fluctuations and cumulants of sigma which are expected to follow the trend of net-proton number cumulants. We emphasize the role of a nonequilibrium first-order phase transition: The presence of an unstable phase causes the well-known bending of the trajectories in the space of temperature and baryochemical potential. For these cases at lower beam energies, the system crosses the freeze-out line more than once, allowing us to calculate a range of cumulants for each initial condition which are overall enhanced for the second hit of the freeze-out line. We thus find not only the critical end point but also the phase transition of the underlying model clearly reflected in the cumulants. The impact of volume fluctuations is demonstrated to play a measurable role for fluid dynamical evolutions that last significantly long. 
\end{abstract}

\vspace{2pc}

\submitto{\PS}
\maketitle

\section{Introduction}
\label{sec:introduction}

Nuclear matter at high temperatures undergoes a transition from a hadronic phase to a quark-gluon plasma (QGP) as demonstrated by experiments at SPS \cite{Heinz:2000bk} and STAR \cite{STAR:2005gfr}. The QGP is characterized as a state of deconfinement and chiral symmetry restoration. Lattice quantum chromodynamics (QCD) techniques have revealed that the transition from hadron gas to QGP is a continuous crossover for zero and small baryochemical potential $\mu_{\rm B}$ \cite{Aoki:2006we,Borsanyi:2010bp,Bazavov:2014pvz}, but so far are not able to give reliable predictions for the region of large $\mu_{\rm B}$ in the QCD phase diagram due to the infamous sign problem. Nevertheless, many studies predict a critical end point (CEP) and first-order phase transition (FOPT). These studies to rely on a variety of effective models of QCD \cite{Scavenius:2000qd,Schaefer:2004en,Fukushima:2008wg} as well as functional techniques \cite{Fischer:2014ata,Gao:2020fbl} which, however, yield widely different results concerning the existence and location of CEP and FOPT in the plane of $T$ and $\mu_{\rm B}$. 

Besides these theoretical approaches, considerable effort is invested into ongoing experimental programs aiming at understanding the QCD phase diagram, e.g. the beam energy scan program at STAR \cite{STAR:2021iop}, NA49/61 \cite{Grebieszkow:2009jr,Andronov:2018ccl}, HADES \cite{HADES:2020wpc} or the upcoming facilities NICA \cite{nica:whitepaper} and FAIR \cite{Friman:2011zz} which are going to focus on low to intermediate beam energies and the physics of compressed baryonic matter. The most prominent observables for studying the presence of a CEP are cumulants of conserved quantities, namely baryon number, strangeness and electric charge. As shown in lattice QCD \cite{Karsch:2016yzt,Bazavov:2020bjn}, various studies of effective models \cite{Skokov:2010uh,Almasi:2017bhq,Wen:2018nkn}, and functional techniques \cite{Isserstedt:2019pgx}, these cumulants exhibit divergences at the CEP and characteristic behavior in the critical region around the CEP including local maxima, minima, and sign changes. Even though all of these calculations are based on equilibrium thermodynamics, it is widely believed that a non-monotonic behavior of cumulants or cumulant ratios as function of center-of-mass energy can be considered a smoking-gun signal for the presence of a CEP. Here, however, a thorough understanding of the nonequilibrium dynamics and design of proper experimental methods and corrections are crucial. For an overview of CEP physics at STAR and the related challenges, see \cite{Luo:2017faz}. 

Cumulants of various order are directly proportional to some power of the correlation length and therefore diverge at the CEP in an infinitely large and equilibrated medium. In a heavy-ion collision, the growth is limited not only by the finite system size but also by critical slowing down \cite{Berdnikov:1999ph}. This, together with finite time effects, and memory effects can lead to a broadening or distortion of the critical region as demonstrated by a variety of nonequilibrium models \cite{Nahrgang:2011mv,Herold:2013bi,Mukherjee:2015swa,Jiang:2015hri,Herold:2017day,Stephanov:2017ghc,Stephanov:2017wlw,Nahrgang:2018afz,Nahrgang:2020yxm}. The impact of these effects on experimental observables, e.g. net-proton number fluctuations, has been studied and reported in \cite{Athanasiou:2010kw,Stephanov:2011pb,Jiang:2015cnt,Herold:2016uvv}. Equally important as understanding the critical dynamics near a chiral CEP is the understanding and modeling of the dynamics at a FOPT where spinodal decomposition enforces density inhomogeneities within single events and also enhances cumulants  \cite{Randrup:2009gp,Randrup:2010ax,Steinheimer:2012gc,Herold:2013qda,Herold:2014zoa,Jiang:2017fas} and is expected to result in an increased production of entropy \cite{Csernai:1992as,Herold:2018ptm}.

In the present paper, we use the nonequilibrium chiral fluid dynamics model \cite{Nahrgang:2011mg} applied to a Bjorken expansion in the longitudinal direction of the beam axis \cite{Herold:2018ptm}. This model describes the dynamics of the sigma field as the chiral order parameter with a Langevin equation interacting with a locally thermalized expanding quark fluid and has proven capable of describing nonequilibrium effects at a CEP and FOPT. Field and fluid are permitted to exchange energy-momentum via a source term. We extract cumulants of the sigma field on an event-by-event basis along a parametrized freeze-out curve. As shown in \cite{Stephanov:2011pb}, it is possible to relate these cumulants to net-proton number cumulants by assuming a superposition of standard Poisson and critical fluctuations. Volume fluctuations are taken into account as described in \cite{Skokov:2012ds}. 

After a description of the model in Section~\ref{sec:model} together with the freeze-out prescription and the calculation method for fluctuations in the chiral order parameter, we present our results on the sigma cumulants as function of beam energy in Section \ref{sec:results}. We  consider the impact of volume fluctuations which we also quantify and investigate further in Section \ref{sec:vol}. Finally, we conclude with a summary and outlook in Section \ref{sec:summary}.

\section{Model description}
\label{sec:model}

Our model provides a dynamical description of the widely studied quark-meson model \cite{Scavenius:2000qd,Mocsy:2004ab,Schaefer:2004en} which exhibits a CEP around $(T_{\rm CP}, \mu_{\rm CP})=(100,200)$~MeV. Here, and in the following equations, $\mu=\mu_{\rm B}/3$ denotes the quark chemical potential. The Lagrangian which includes light quarks $q=(u,d)$ and the chiral order parameter $\sigma$ reads
\begin{align}
\label{eq:Lagrangian}
 {\cal L}&=\overline{q}\left(i \gamma^\mu \partial_\mu-g \sigma\right)q + \frac{1}{2}\left(\partial_\mu\sigma\right)^2- U(\sigma)~, \\
 U(\sigma)&=\frac{\lambda^2}{4}\left(\sigma^2-f_{\pi}^2\right)^2-f_{\pi}m_{\pi}^2\sigma +U_0~,    
\end{align}
with standard parameters $f_\pi=93$~MeV, $m_\pi=138$~MeV and $U_0$ such that the potential is equal to zero in the ground state. Here, the value of the pion fields has already been fixed at its vacuum expectation value of zero as these are irrelevant for the phase structure of the model and the dynamics of the nonequilibrium phase transition. The quark-sigma coupling constant $g$ is determined from the requirement that $g\sigma$ equals the nucleon mass of $940$~MeV in vacuum. 

The grand potential $\Omega$ in mean-field approximation reveals the phase structure of the quark-meson model and is evaluated as $\Omega=\Omega_{q\bar q}+U$, with the quark-antiquark contribution
\begin{align}
 \Omega_{q\bar q}=-2N_f N_c T\int \frac{\mathrm d^3 p}{(2\pi)^3} & \left[\log\left(1+\mathrm e^{-\frac{E-\mu}{T}}\right)\right. \\ & \left. +\log\left(1+\mathrm e^{-\frac{E+\mu}{T}}\right)\right]~.\nonumber
\end{align}
Here, $N_f=2$, $N_c=3$, represent the number of light quark flavors and colors, respectively, and $E=\sqrt{p^2+g^2\sigma^2}$ denotes the quark quasiparticle energy, assuming a dynamically generated quark mass of $m\sigma$.

\subsection{Equations of motion}

In the present work, we focus on the fluctuations of the sigma field or, more precisely, the zero mode or volume-average of the sigma field which at any given time is defined as 
\begin{equation}
    \sigma=\frac{1}{V}\int\mathrm d^3 x\sigma(x)~.
\end{equation}
This quantity is widely regarded as the critical mode of the QCD CEP and characterized by a vanishing mass and diverging correlations length at the CEP \cite{Stephanov:2008qz,Stephanov:2011pb}.

Since we describe the expanding fluid using a Bjorken model, we use proper time $\tau$ rather than coordinate time $t$, and the dots in the following equations denote derivatives with respect to $\tau$. We evolve the zero mode of the sigma field using a Langevin equation of motion derived from the two-particle irreducible effective action \cite{Nahrgang:2011mg}, 
\begin{equation}
 \label{eq:eom_sigma}
 \ddot\sigma+\left(\frac{D}{\tau}+\eta\right)\dot\sigma+\frac{\delta\Omega}{\delta\sigma}=\xi~.
\end{equation}
Here, we neglect any spatial fluctuations. For our case of purely longitudinal hydrodynamic flow, we set $D=1$ in the Hubble term. The full and proper nonequilibrium dynamics of sigma is encoded in the dissipation coefficient $\eta$ and the stochastic noise $\xi$ which are related by a dissipation-fluctuation relation, 
\begin{equation}
 \langle\xi(t)\xi(t')\rangle=\frac{m_{\sigma}\eta}{V}\coth{\left(\frac{m_{\sigma}}{2T}\right)}\delta(t-t')~.
\end{equation}
Here, $\xi$ is assumed Gaussian with mean zero and white, i.e. it is not correlated over time. The screening mass of the sigma field $m_{\sigma}$ is given by the second derivative of $\Omega$ with respect to $\sigma$ evaluated at equilibrium, $\sigma=\bar \sigma$,
\begin{equation}
    m_{\sigma}=\left.\frac{\partial \Omega}{\partial\sigma}\right|_{\sigma=\bar \sigma}~.
\end{equation}

The damping coefficient $\eta$ includes effects from various processes, such as:
\begin{itemize}
    \item Mesonic interactions, i.e.\ scattering of a condensed sigma meson with a thermal sigma, $\sigma\sigma\leftrightarrow\sigma\sigma$, and  $\sigma\leftrightarrow\pi\pi$ (two-pion decay) \cite{Csernai:1999ca}, described by a phenomenological damping coefficient of $\eta=2.2/$fm \cite{Biro:1997va} wherever kinematically allowed. 
    \item Meson-quark interactions, $\sigma\leftrightarrow q\bar q$, leading to a $T$- and $\mu$-dependent coefficient \cite{Nahrgang:2011mg}, 
    \begin{equation}
        \eta=\frac{12 g^2}{\pi}\left[1-2n_{\rm F}\left(\frac{m_\sigma}{2}\right)\right]\frac{1}{m_\sigma^2}\left(\frac{m_\sigma^2}{4}-m_q^2\right)^{3/2}~.
    \end{equation}
\end{itemize}

We assume an ideal fluid of quarks and antiquarks described by the energy-momentum tensor $T^{\mu\nu}_q=(e+p)u^{\mu}u^{\nu}-pg^{\mu\nu}$. Due to energy-momentum conservation the divergence of the total energy-momentum tensor $T^{\mu\nu}_q+T^{\mu\nu}_\sigma$ vanishes, which leads to 
\begin{equation}
\label{eq:eom_fluid}
 \partial_\mu T^{\mu\nu}_q=\left[\frac{\delta\Omega_{q \bar q}}{\delta\sigma}+\left(\frac{D}{\tau}+\eta\right)\dot\sigma\right]\partial^\nu \sigma.
\end{equation}
A contraction of \eqref{eq:eom_fluid} with the four-velocity $u^\nu$ yields the evolution equation for the energy density,
\begin{equation}
\label{eq:eom_eden}
 \dot e=-\frac{e+p}{\tau}+\left[\frac{\delta\Omega_{q \bar q}}{\delta\sigma}+\left(\frac{D}{\tau}+\eta\right)\dot\sigma\right]\dot\sigma~,
\end{equation}
while the net-baryon density simply follows
\begin{equation}
 \label{eq:eom_nden}
 \dot n = -\frac{n}{\tau}~.
\end{equation}
In equations \eqref{eq:eom_sigma} and \eqref{eq:eom_eden} the pressure is given by $p=-\Omega_{q\bar q}$. It is therefore an explicit function not only of $T$ and $\mu$, but also of $\sigma$ which during the evolution is not fixed to its equilibrium value. 

We solve the sigma equation \eqref{eq:eom_sigma} numerically using a method for Langevin equations \cite{CassolSeewald:2007ru} and equations \eqref{eq:eom_eden}, \eqref{eq:eom_nden} for $e$, $n$ with a simple Runge-Kutta algorithm. For the numerical implementation of e.g. the stochastic fluctuations and the study of volume fluctuations as it will presented in Section \ref{sec:vol}, a proper definition of the fireball volume $V$ is imperative. We set 
\begin{equation}
\label{eq:volume}
    V=\pi R^2 \tau
\end{equation}
according to the dynamics of the Bjorken model with a cylindrical volume of cross section $\pi R^2$ representing a central collision of two gold nuclei with radii $R=7.3$~fm.

\subsection{Freeze-out and mapping to beam energies}

We investigate cumulants of the chiral field up to order four which are going to be extracted along a parametrized freeze-out curve. The quark-meson model has a phase diagram which we do not expect to resemble real QCD, e.g. the crossover temperature is significantly lower than predicted by lattice QCD data, and the FOPT at low temperatures is at unrealistically small values of the chemical potential. Therefore, we have to restrict ourselves to a qualitative description of the freeze-out and do this by using a simple parametrized freeze-out curve that has been obtained from thermal model fits to Au+Au and Pb+Pb collision over an energy range from $\sqrt{s}=2.24$ to $200$~AGeV (SIS, AGS, SPS, RHIC) \cite{Cleymans:2005xv}. The proposed parametrization of the curve in this publication reads
\begin{equation}
    \label{eq:freeze}
    T_{\rm f.o.}(\mu_B) = a - b\mu_B^2 - c\mu_B^4~,
\end{equation}
with constants $a=0.166$~GeV, $b=0.139$~GeV$^{-1}$, and  $c=0.053$~GeV$^{-3}$. Since the phase boundary of the quark-meson model which presents the underlying phase diagram of our work, would lie below the thus obtained freeze-out line, we scale both $T$ and $\mu$ in this parametrization with a common factor $T_{\rm crossover}(\mu=0)/T_{\rm f.o.}(\mu=0)$. This results in a scaled-down freeze-out curve which corresponds with the crossover temperature at $\mu=0$ and consistently lies below the crossover and phase transition for positive values of $\mu$. Here, the crossover temperature of the quark-meson model,  $T_{\rm crossover}(\mu=0)=145$~MeV, has been determined from a maximum of the quark number susceptibility.  

We define initial conditions for the evolution by choosing initial values $T_{\rm i}$ and $\mu_{\rm i}$ similar to what was used in previous studies \cite{Herold:2018ptm}. During the evolution of the fluid according to equations~\eqref{eq:eom_sigma}, \eqref{eq:eom_eden}, \eqref{eq:eom_nden}, the trajectory in $T$-$\mu$ space will hit the freeze-out curve. This hit point is then used to map the evolution to a corresponding beam energy via
    \begin{equation}
    \label{eq:matchsqrts}
    \mu_B(\sqrt{s}) = \frac{d}{1 + e\sqrt{s}}~,
    \end{equation}
with parameters $d = 1.308$ GeV and  $e = 0.273$ GeV$^{-1}$ \cite{Cleymans:2005xv}. Since the hit point fluctuates from event to event due to the stochastic nature of the evolution, we use event-averaged values for $\mu$ in equation~\eqref{eq:matchsqrts}.

\subsection{Sigma cumulants and corrections for volume fluctuations}

As argued earlier \cite{Stephanov:2011pb}, one can derive a direct relation between cumulants in the zero mode of the sigma field $\sigma_V$ and the experimentally accessible net-proton number by assuming a suitable coupling between sigma and protons $g\sigma \bar p p$. This leads to a fluctuation of the particle multiplicity $N$ as
\begin{equation}
    \delta N = \delta N_0 + g \delta \sigma_V d \int \frac{\mathrm d^3 p}{(2\pi)^3}\frac{\partial n}{\partial m}~.
\end{equation}
In this equation, $\delta N_0$ denotes the fluctuation according to a Poisson or Skellam distribution, $d$ the degeneracy factor (equal to $2$ for protons) and $n$ the equilibrium distribution function which is a function of the fluctuating  particle mass $m=g\sigma$. As a consequence of this, any enhancement or suppression of cumulants of $\delta N$ is going to be proportional to the respective cumulant of $\delta \sigma_V$.

Here, we report results for cumulants of $\sigma_V=\int \mathrm d^3 x \sigma = \sigma V$, equal to the product of the sigma field and the volume at freeze-out since our approach does not include the description of spatial fluctuations. We calculate the cumulants $C^{\sigma}_n$ of $\sigma_V$, where $\langle \cdot \rangle$ denotes an averaging over events. The explicit formulas for $n=1, 2, 3, 4$ read
\begin{align}
    C^{\sigma}_1&=\langle \sigma_V\rangle~, \\
    C^{\sigma}_2&=\langle \delta\sigma_V^2\rangle~, \\
    C^{\sigma}_3&=\langle \delta\sigma_V^3\rangle~, \\
    C^{\sigma}_4&=\langle \delta\sigma_V^4\rangle - 3\langle\delta\sigma_V^2\rangle^2~
\end{align}
where $\delta\sigma_V=\sigma_V-\langle\sigma_V\rangle$. The cumulants are therefore identical to the corresponding central moments for $n=2,3$ and differ by an additional term for $n=4$.

To account for a fluctuating freeze-out volume, we impose the corrections proposed in an earlier work \cite{Skokov:2012ds}. Considering reduced cumulants $c^{\sigma}_n=C^{\sigma}_n/\langle V \rangle$, the corresponding corrected quantities read
\begin{align}
    c^{\sigma,\mathrm{corr}}_1=&c^{\sigma}_1 \label{eq:volcorr1}\\
    c^{\sigma,\mathrm{corr}}_2=&c^{\sigma}_2-(c^{\sigma,\mathrm{corr}}_1)^2 v_2~, \label{eq:volcorr2} \\
    c^{\sigma,\mathrm{corr}}_3=&c^{\sigma}_3-3(c^{\sigma,\mathrm{corr}}_2)(c^{\sigma,\mathrm{corr}}_1)v_2\label{eq:volcorr3}\\&-(c^{\sigma,\mathrm{corr}}_1)^3 v_3~, \nonumber   \\
    c^{\sigma,\mathrm{corr}}_4=&c^{\sigma}_4-\left[4(c^{\sigma,\mathrm{corr}}_3)  (c^{\sigma,\mathrm{corr}}_1)+3(c^{\sigma,\mathrm{corr}}_2)^2\right]v_2\label{eq:volcorr4} \\&-6(c^{\sigma,\mathrm{corr}}_2)(c^{\sigma,\mathrm{corr}}_1)^2 v_3-(c^{\sigma,\mathrm{corr}}_1)^4 v_4~. \nonumber
\end{align}
In this notation, the $v_n$ are reduced cumulants of the volume fluctuations, i.e. $v_n=\langle\delta V^n\rangle/\langle V\rangle$. From $n=2$ on, the calculation of the corrected cumulant requires knowledge of all orders of cumulants up to $n-1$.

\section{Research procedure and results}
\label{sec:results}

\begin{figure}[tbp]
\centering
\includegraphics[width=.65\textwidth,origin=c]{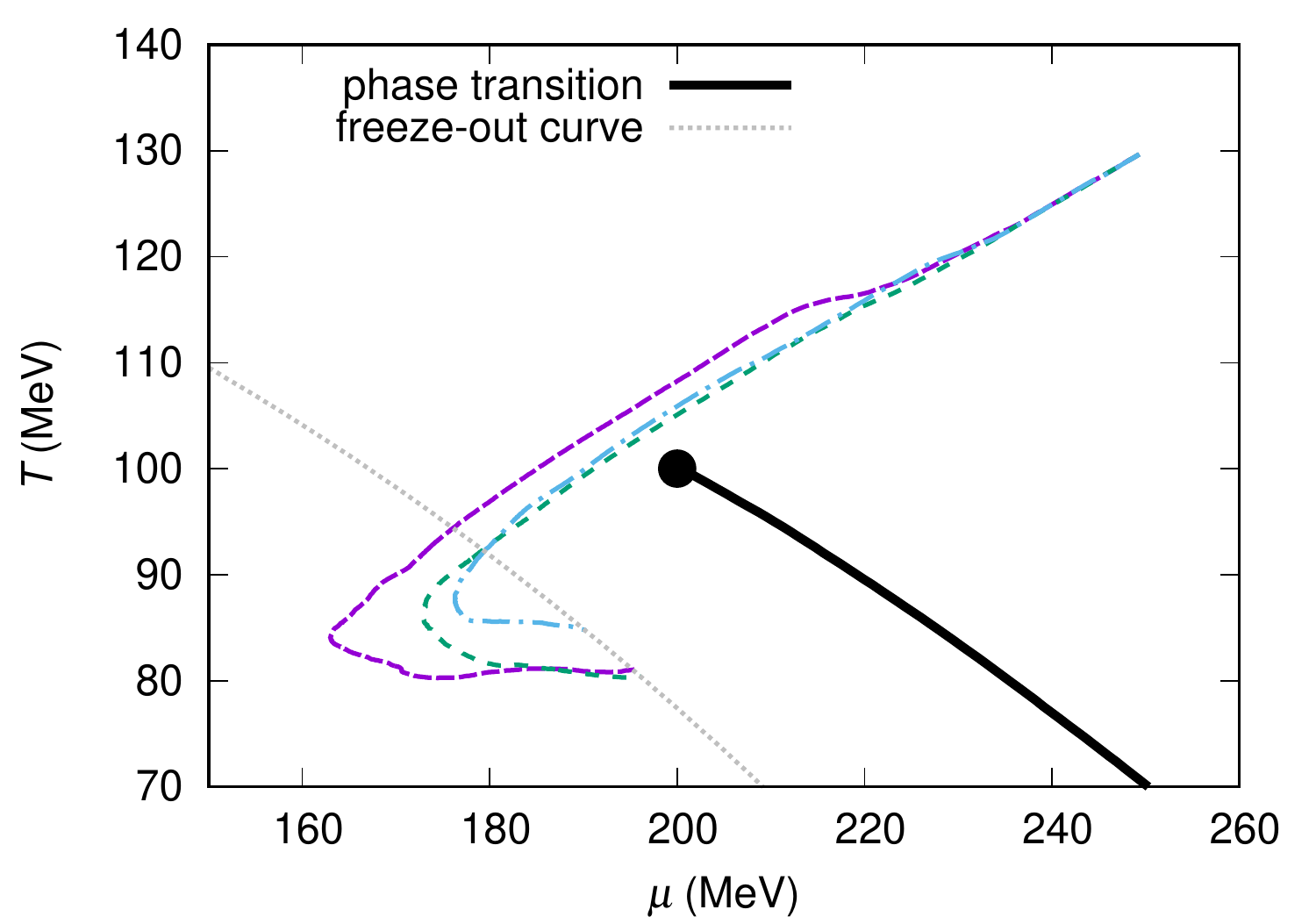}
\caption{Three exemplary trajectories from the same initial condition passing near the CEP. The back-bending after crossing the critical region leads to more than one hit with the freeze-out curve for each trajectory.}
\label{fig:traj}
\end{figure}

\begin{figure}[tbp]
\centering
\includegraphics[width=.65\textwidth,origin=c]{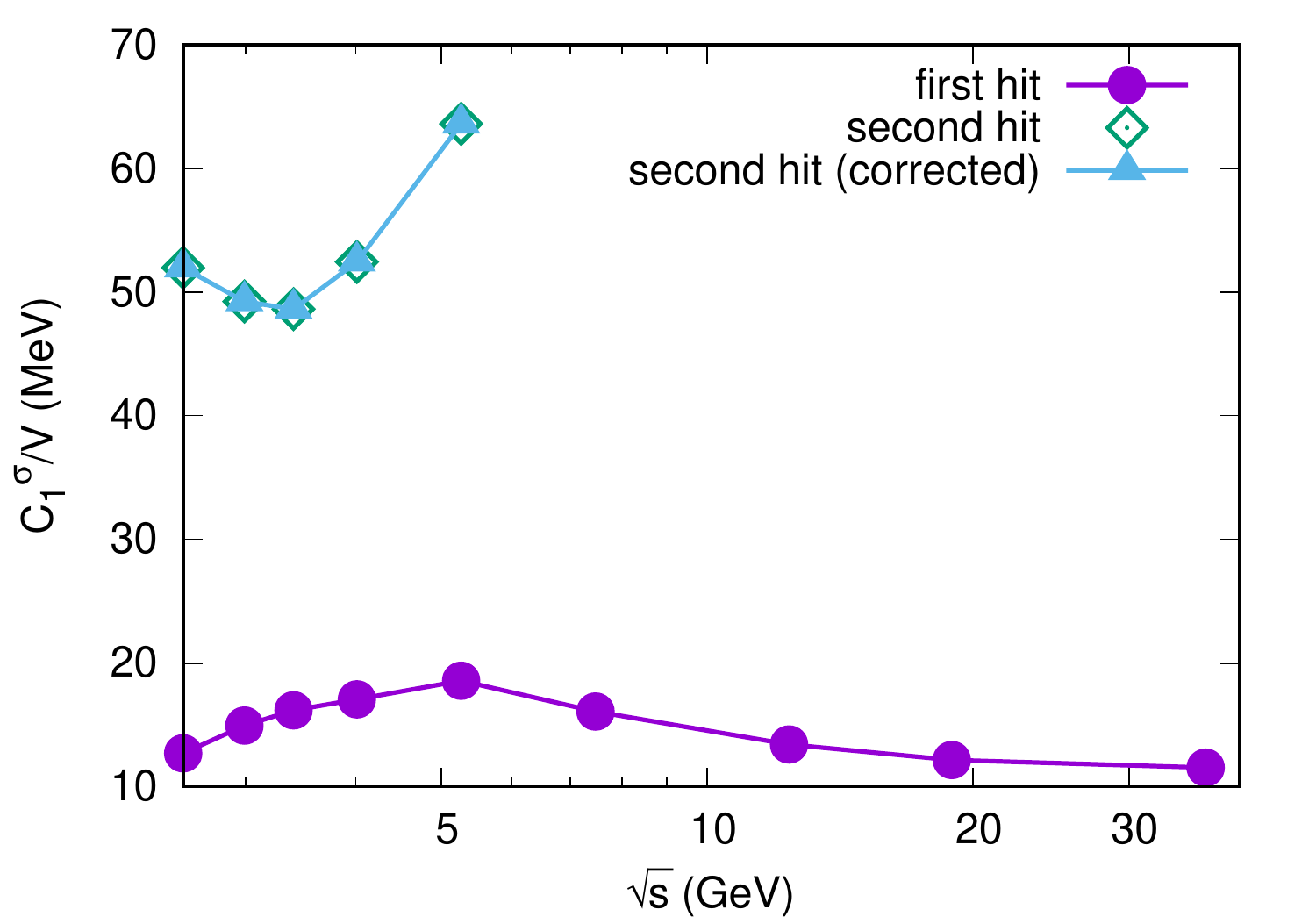}
\caption{First cumulant of the sigma field scaled by volume as function of beam energy, for first and second hit of the freeze-out curve.}
\label{fig:c1sig}
\end{figure}

We initialize the fluid at a set of fixed values ($T_i$,$\mu_i$) borrowed from previous work \cite{Herold:2018ptm} that represent various initial conditions intended to probe crossover, CEP, and FOPT regions in the chiral phase diagram. We then calculate the initial values of the sigma field, energy density, quark number density, and pressure as the corresponding equilibrium values. We let the coupled system of field and fluid evolve according to the equations of motion \eqref{eq:eom_sigma}, \eqref{eq:eom_eden}, \eqref{eq:eom_nden} until the freeze-out curve is hit. The parametrized freeze-out curve defines the point at which we calculate physical observables for a given beam energy. Note, however, that the choice of this freeze-out condition is as usual somewhat arbitrary and other conditions might be conceived. Notably, for expansions at high baryochemical potential, the freeze-out curve can be hit more than once due to the nonequilibrium evolution of the expanding plasma which results in a back-bending of the trajectories in the plane of $T$ and $\mu$. The effect is related to a release of latent heat and significant energy dissipation that drives the system back into the direction of the chirally restored phase. A similar, albeit less pronounced effect also occurs for higher beam energies or lower initial values of $\mu$. Even though there is no phase transition, the chiral order parameter experiences a delayed relaxation. Consequently, the delayed release of latent heat leads to a similar back-bending effect visible in the trajectories \cite{Herold:2018ptm}. In contrast to that, an ideal hydrodynamic system without source term and stochastic fluctuations would preserve the  entropy-per-baryon $S/A$ and therefore evolve along the isentropes of the underlying model. This would lead the system along the phase boundary for a certain amount of time \cite{Steinheimer:2007iy}. To take into account the effect of a possible mixed phase in our nonequilibrium model, we evolve the system with sufficiently large initial $\mu_i$ until a second crossing of the freeze-out curve. While the system evolves through the spinodal region, it bends back and has the chance to hit the freeze-out condition another time. One can assume that between first and second hit, the system has not fully relaxed into the chirally broken phase but is undergoing droplet formation leading to a gradual freeze-out \cite{Mishustin:1998eq,Randrup:2009gp}. Effectively, this prolongs the freeze-out process and we take this into account by defining two boundaries for which we calculate the corresponding cumulants. Thus, the cumulants at these two boundaries or hit points may be understood as a range of possible values  for the cumulants after freeze-out. Subsequently, we calculate cumulants at both hit points separately, providing us with a range of possible values for these scenarios at lower beam energies. This plays an important role for evolutions crossing the FOPT line or near the CEP, see figure \ref{fig:traj} for some examples of possible trajectories. Numerically, the stochastic noise term $\xi$ is randomly generated in each time step from a corresponding Gaussian distribution, making the evolution non-deterministic. This is in turn influencing the energy density in equation \eqref{eq:eom_eden} and finally the evolution of temperature and chemical potential. The figure shows three trajectories starting from the same initial condition and evolving slightly differently due to the presence of stochastic fluctuations in the equation of motion \eqref{eq:eom_sigma}. Consequently, the freeze-out points differ between these. For matching with $\sqrt{s}$ according to equation~\eqref{eq:matchsqrts}, we only use the first hit point and determine the event-averaged values of $T$ and $\mu$ at this final point to calculate the center-of-mass energy $\sqrt{s}$ corresponding to this initial condition according to equation~\eqref{eq:matchsqrts}. 

We simulate $N=10^7$ events and calculate event-by-event fluctuations in terms of cumulants of $\sigma_V$. Since these are subject to significant fluctuations of the freeze-out volume, we also determine the corresponding necessary corrections according to equations \eqref{eq:volcorr1}-\eqref{eq:volcorr4}. We expect that these corrections play an important role for second hits of the freeze-out curve which take an extended amount of time and therefore result in a larger range of freeze-out times and, consequently, freeze-out volumes, cf. equation~\eqref{eq:volume}. In the figures and results below we show the different cumulants with and without volume corrections only for these cases since the differences for the first hits of the freeze-out line turned out to be insignificant altogether. 

Figure \ref{fig:c1sig} shows the first cumulant or average of $\sigma_V$ scaled by the average volume $V$. We see that the obtained values for the first hits range between $10-20$~MeV, and significantly above that for the second hits. This is easily understood considering that for the latter cases, the system evolves over a longer time, therefore allowing for further relaxation and approach to the low-temperature equilibrium state which lies around $f_{\pi}=93$~MeV in the chirally broken phase, while for the first hits they still remain close to their equilibrium value in the chirally restored phase, which is slightly above zero. 

\begin{figure}[tbp]
\centering
\includegraphics[width=.65\textwidth,origin=c]{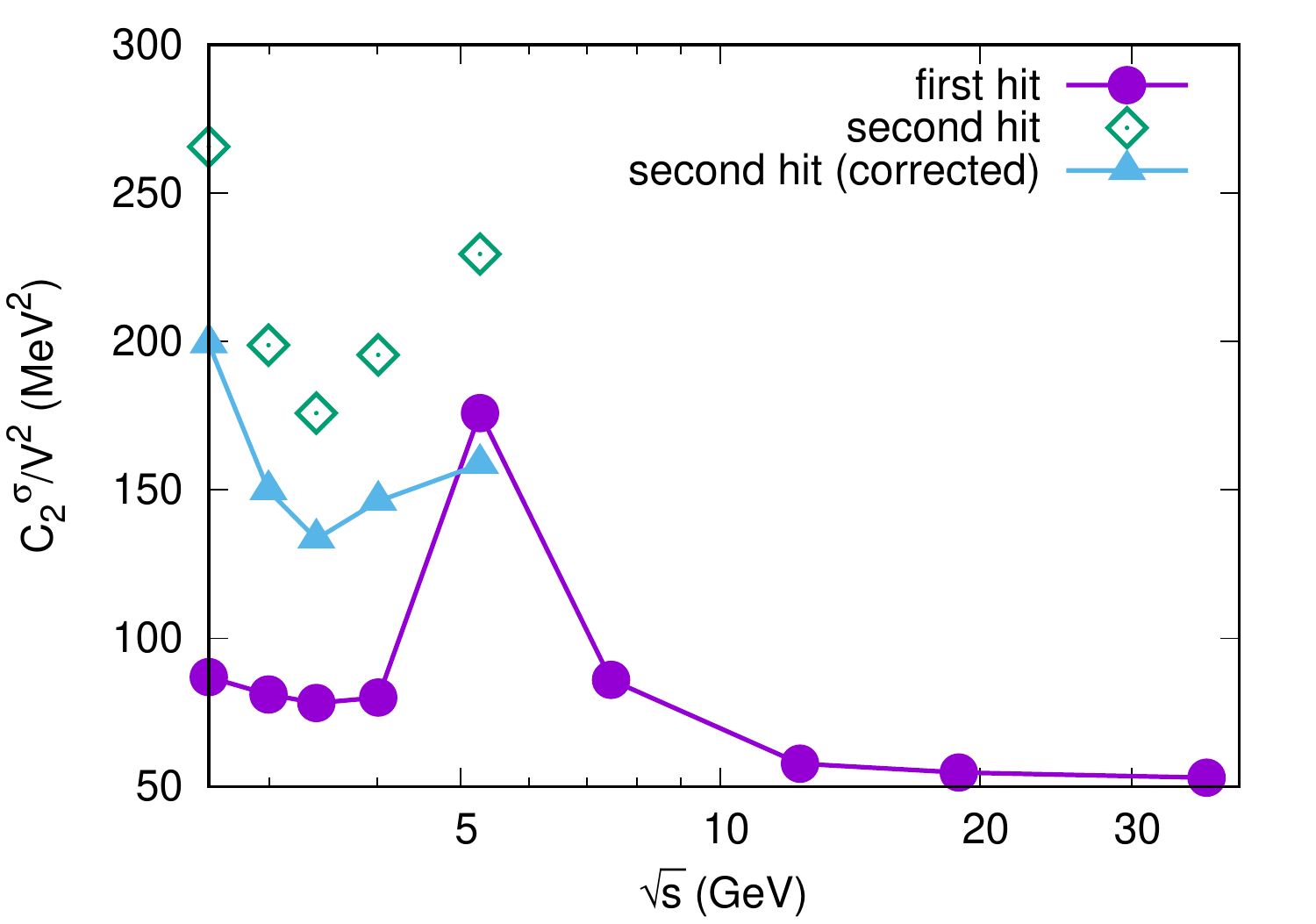}
\hfill
\includegraphics[width=.65\textwidth,origin=c]{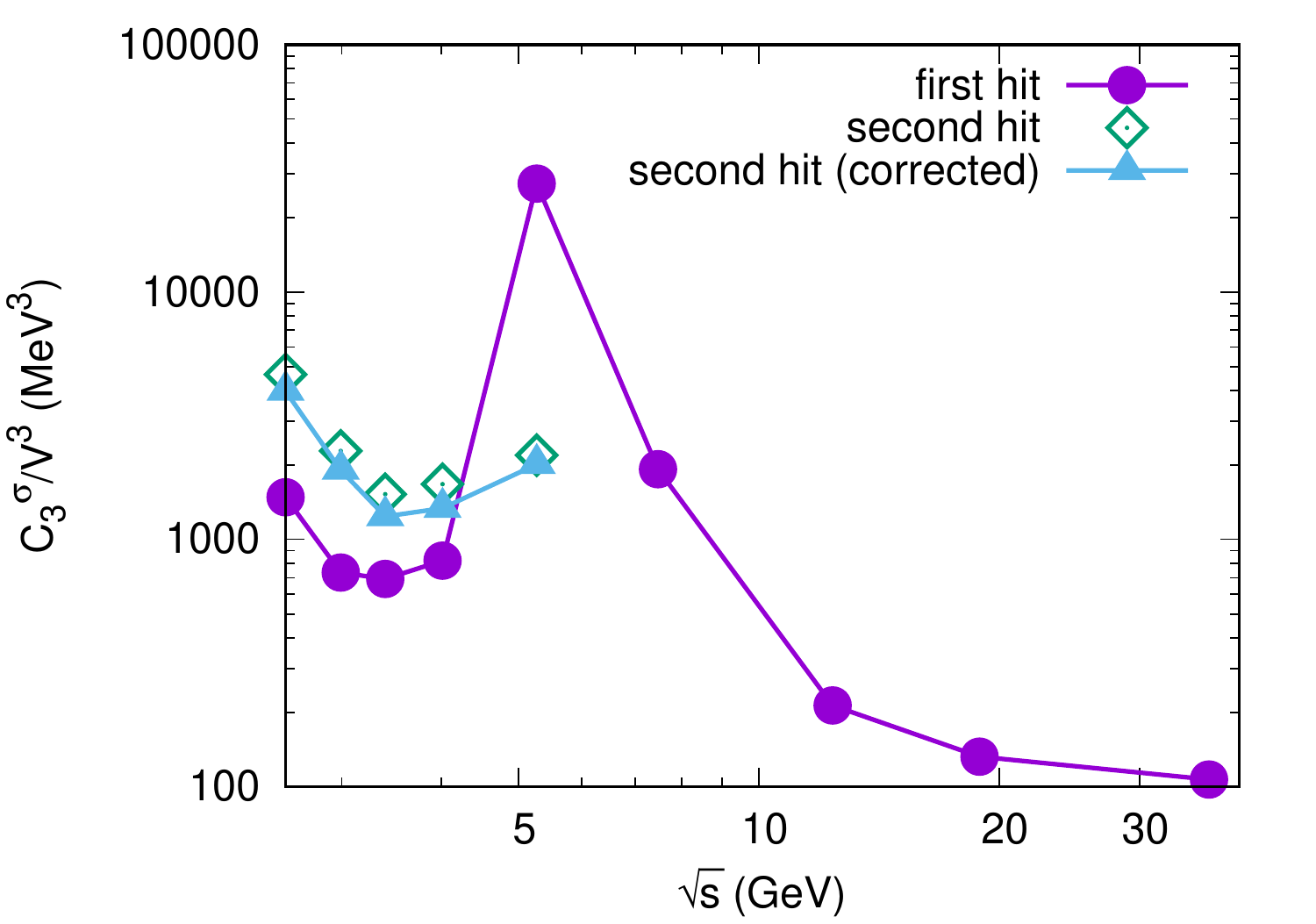}
\hfill
\includegraphics[width=.65\textwidth,origin=c]{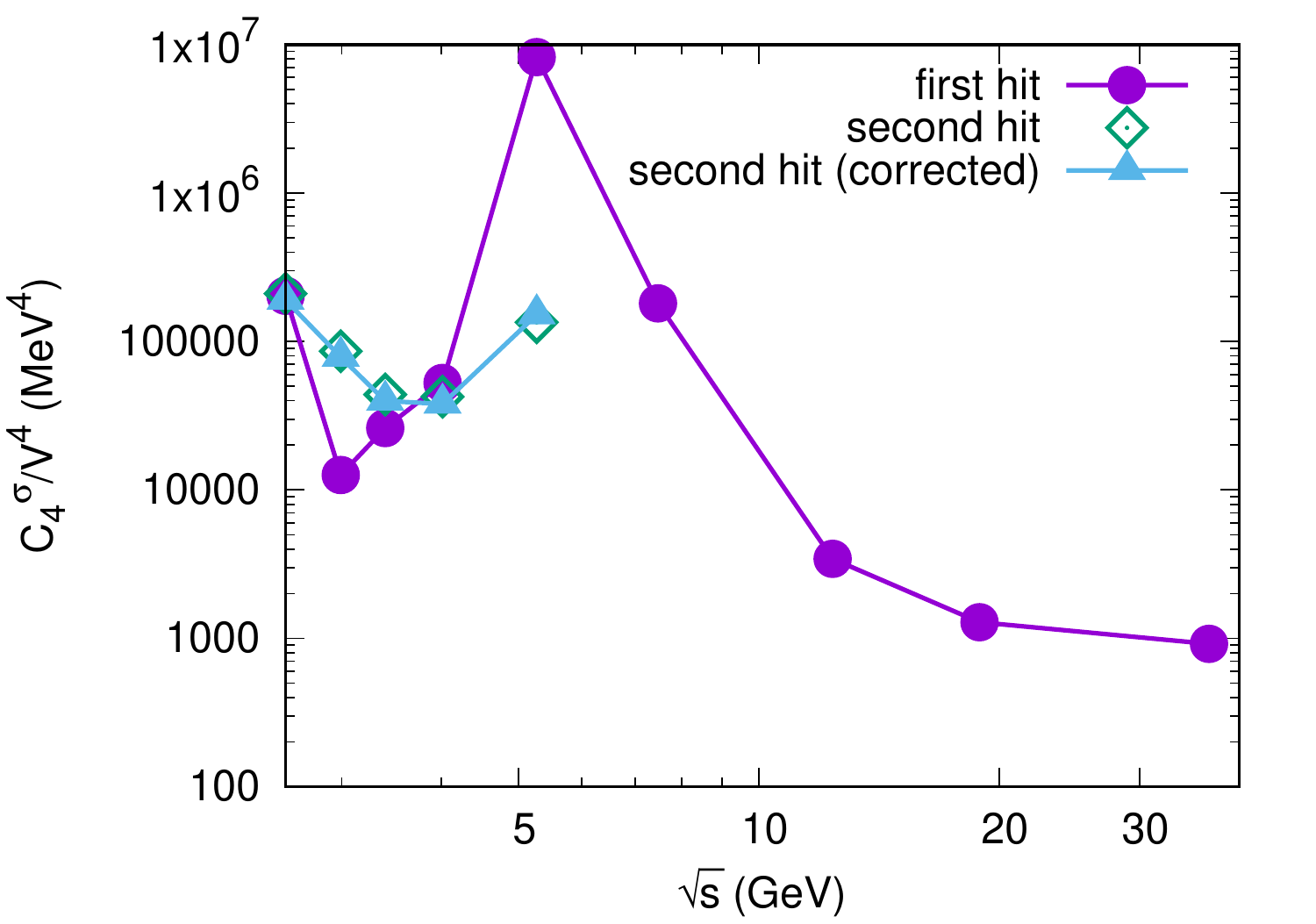}
\caption{Second (top), third (middle), and fourth (bottom) scaled cumulant of the sigma field as function of beam energy, for first and second hit of the freeze-out curve. For the second hits, a comparison to corrections for volume fluctuations is shown.}
\label{fig:c234sigma}
\end{figure}

Figure~\ref{fig:c234sigma} shows our obtained results for the second-, third-, and fourth-order cumulants, scaled by the respective power of freeze-out volume $V^n$. All figures show that the volume corrections decrease the obtained values at the second hit by around $20-30\%$. Note here that the middle and bottom figures with $n=3,4$ use logarithmic axes for the cumulants $C^\sigma_3$ and $C^\sigma_4$. The second-order cumulant in the upper figure increases with decreasing beam energy until it reaches a maximum at around $5$~GeV for the evolution that passes through or very close to the CEP and freezes out precariously close to a spinodal line. As shown in \cite{Sasaki:2007qh,Herold:2014zoa}, the spinodal lines around the FOPT exhibit divergences of all susceptibilities in nonequilibrium when properly taking spinodal instabilities into account. A careful analysis of the divergences by calculating the critical exponents reveals that they become stronger with increasing order $n$ of the cumulant \cite{Sasaki:2007qh,Herold:2014zoa}. The divergence is clearly reflected in the peak in $C^{\sigma}_2/V^2$ and furthermore in the peaks visible in the third- and fourth-order cumulants at the same energy. Going to lower beam energies, the second-order cumulant decreases again but overall remains at a higher level than at the larger beam energies. A significant enhancement is found when comparing to the cumulants at the second hit points of the freeze-out curve. These surpass those of the first hits in all cases except for the evolution with  initial energy around $5$~GeV, most likely as a result of a further enhancement of fluctuations as the system passes through the mixed-phase region. 

The middle and bottom plots in figure~\ref{fig:c234sigma} depict results for the third- and fourth-order cumulants, again scaled by the respective power of $V$. Just as with the second-order cumulant, both exhibit a clear peak at $5$~GeV which is not surpassed by the cumulants at the second hit. For lower energies, they are again in generally lying above the cumulants from the first hit of the freeze-out line. 

In all three subplots of figure~\ref{fig:c234sigma} we another common trend: An local maximum at the lowest beam energy or an increase of the respective cumulant when approaching the lowest beam energy which becomes more significant with increasing order of the cumulant $n$. This can be understood as an indicator for the potential of signals at the FOPT that might grow similarly strong or even stronger than those of the CEP. Assuming the behavior of the net-proton number cumulants follows the behavior of the sigma cumulants, we may expect them to be significantly enhanced above the Poisson baseline for evolutions through CEP and FOPT. For future experiments that explore the lowest beam energy and highest densities, the onset of the FOPT would thus be clearly visible by a sudden rise of the cumulant.

For the determination of volume-independent cumulant ratios of baryon number, electric charge, and strangeness, which are the focus of the related experiments, further detailed studies and calculations are necessary and currently in progress. Considering the energy range covered by STAR (above $7.7$~GeV) and HADES ($2.4$~GeV), we finally note that coverage of the energies between these is desirable as in this range, the peak resulting from the critical and/or spinodal region is found.

\section{Volume fluctuations}
\label{sec:vol}

\begin{figure}[tbp]
\centering
\includegraphics[width=.65\textwidth,origin=c]{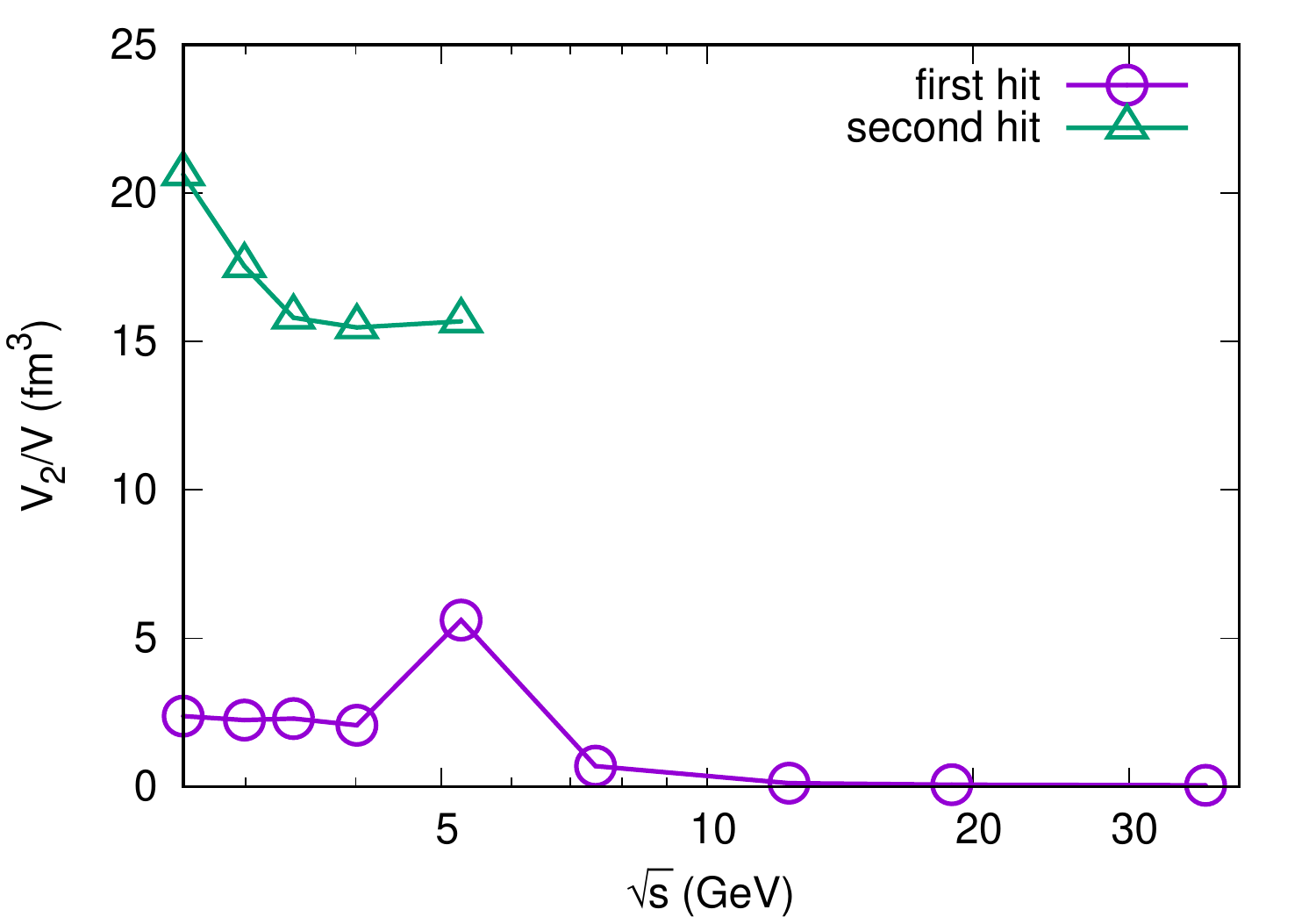}
\caption{Fluctuations of the volume as function of beam energy, for first and second hit of the freeze-out curve.}
\label{fig:volfluct}
\end{figure}

We investigate the volume fluctuations quantitatively to understand their impact and confirm our assumptions from the previous section. To do so, we calculate the variance of the volume $V_2$ as
\begin{equation}
 V_2 = \langle \delta V^2\rangle~,~~ \delta V = V-\langle V\rangle~.
\end{equation}
 Figure~\ref{fig:volfluct} shows $V_2/V$ as function of $\sqrt{s}$, again for the first hit and the second hit. As suspected, the fluctuations are significantly larger at the second hit compared to the first one, by a factor of $3-10$. This is understood considering the longer evolution and later freeze-out time for these cases, where due to stochastic fluctuations in the evolution equations, a potentially wider range of freeze-out times is obtained. Since the volume scales with the proper time, this also results in a wider range of freeze-out volumes. Interestingly, a slight increase in fluctuations within the set of first hits is found for the evolution through the CEP, whereas much smaller than for the extended evolution and the second hit. 
 
\section{Summary}
\label{sec:summary}
We have studied cumulants of the volume-averaged sigma field $\sigma_V$ up to fourth order within a nonequilibrium chiral Bjorken expansion. A quark-meson model served as input for providing the chiral phase structure and cumulants of the sigma field have been calculated event-by-event at a parametrized freeze-out curve obtained from experimental hadron multiplicity ratios. For low beam energies, we find that the nonequilibrium evolution results in back-bending of the trajectories in the phase diagram and an extended lifetime of the mixed phase. The freeze-out curve is for these cases mostly hit more than once and we regard this behavior by reporting a range of cumulant values for these cases. We found three main results: First, evolutions passing close by or through the CEP of the underlying model result in a maximum of each cumulant as function of beam energy when considering only the first hit of the freeze-out curve. Second, cumulants are enhanced for a FOPT at energies below the CEP evolution, for the second-order cumulant even for the CEP. And finally, the effect of volume corrections is relevant for the cumulants of the CEP and FOPT, where they lead to an increase of about $20-30\%$. The individual evolutions vary more and more in time and therefore expansion volume until they meet the corresponding freeze-out point. 

In the future, we are going to extract cumulants of the net-proton number by either using the direct relation to sigma cumulants or applying a particlization scheme. This will allow direct comparison to results reported by STAR for energies $\sqrt{s_{\rm NN}}\ge 7.7$~GeV \cite{STAR:2021iop} and HADES for $\sqrt{s_{\rm NN}}= 2.4$~GeV \cite{HADES:2020wpc}. Furthermore, an extension to full (3+1) dimensional hydrodynamics is necessary to describe spatial fluctuations.

\ack

This work was supported by (i) Suranaree University of Technology (SUT), (ii) Thailand Science Research and Innovation (TSRI), and (iii) National Science Research and Innovation Fund (NSRF), project no. 160355. This research has received funding support from the NSRF via the Program Management Unit for Human Resources \& Institutional Development, Research and Innovation [grant number B16F640076].

\section*{References}
\bibliographystyle{iopart-num}
\bibliography{mybib.bib}

\end{document}